\documentclass[fleqn]{article}

\usepackage{graphicx}

\begin{document}

\title{Nonrelativistic energies and predissociation widths of quasibound states in the Li$^{3+}\hbox{-}\,d\mu$ molecular ions.}
\author{V.I.~Korobov$^{a,b}$, A.V.~Eskin$^b$, F.A.~Martynenko$^b$, and O.S.~Sukhorukova$^b$\\
\small\sl $^a$Joint Institute for Nuclear Research, Dubna, 141980, Russia\\
\small\sl $^b$Samara National Research University, Samara, 443086, Russia}

\date{}

\maketitle

\begin{abstract}
Muonic molecular ions $^6$Li$^{3+}d\mu$ and $^7$Li$^{3+}d\mu$ are studied numerically. Using the complex coordinate rotation method we found six rotational states (three for each isotope), which are resonant states with the life-time of an order of picoseconds. These molecular systems may be of interest for studying low-energy fusion reactions. A key quantity, $|\Psi(0)|^2$, the wave function squared at the coalescence point of the nuclei is calculated for $S$ states for both isotopes with a precision better than 3\%.
\end{abstract}

\section{Introduction}
Low-energy interactions between atomic nuclei are of great importance, they provide valuable information on the nucleon--nucleon interaction \cite{Friar02}, are used as input data in astrophysics and cosmology \cite{astro1,astro2}, light nuclear reactions are involved in the primordial nucleosynthesis, the $^7$Li abundance observed from halo dwarf stars presents a definite discrepancy with our knowledge on the nuclear rates \cite{Field11}. Very promising physics of fast monoenergetic neutron sources is expected in the laser-driven $d+^7\textrm{Li}$ reactions:
\begin{equation}
d+\,^7\textrm{Li}\to \,^8\textrm{Be}+n+15.03\textrm{ MeV},
\end{equation}
at keV energies \cite{Zhao16}.

Nucleus-nucleus collision data are not much available at energies below the keV region \cite{low}. At the same time nuclear reactions occurring inside the stellar objects proceed dominantly with these low energies. On the other hand when nuclei are confined in a small molecular objects like muonic molecules \cite{Ponomarev90}, the fusion reaction rate may be measured with confident precision. In this case, the rate is proportional to the probability density of two nuclei to be inside the fusion region: $|\Psi(0)|^2$. The latter quantity may be obtained from \emph{ab initio} calculations of the nonrelativistic Schr\"odinger equation for this molecule and then may be used in the experimental studies to extract a proper astrophysical $S$-factor for this reaction.


\begin{figure}
\begin{center}
\includegraphics*[width=0.6\textwidth]{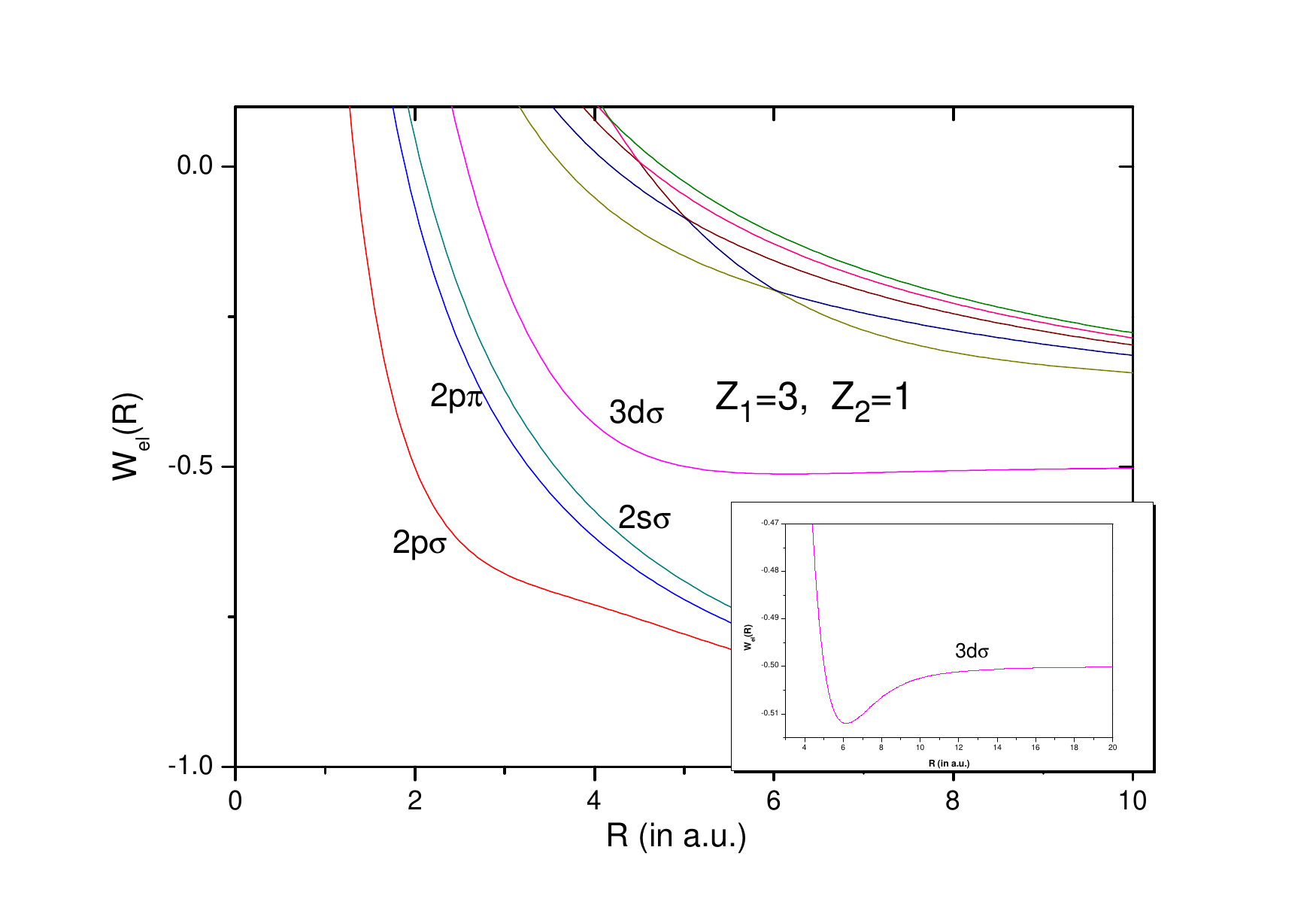}
\end{center}
\caption{Adiabatic potentials, $W_{\rm el}(R)=\mathcal{E}_{\rm el}(R)+Z_1Z_2/R$, for $Z_1\!=\!3$ and $Z_2\!=\!1$. $\mathcal{E}_{\rm el}(R)$ is the electron energy in the two-center Born-Oppenheimer approximation. Inset is the $3d\sigma$ adiabatic potential, which is responsible for existence of quasibound states. The potential curves were calculated by us using the variational approximation based on the spheroidal coordinates for the Coulomb two-center problem.}\label{fig:ad}
\end{figure}

In this paper we want to present precise numerical calculations of the $^6$Li$d\mu$ and $^7$Li$d\mu$ quasibound states, which may be of use for the study of the low energy reactions. Existence of these states may be explained using the adiabatic approximation. Figure \ref{fig:ad} qualitatively illustrates the spectrum of the considered system. The $3d\sigma$ potential in the separated atom limit represents a cluster: \linebreak $d\mu(n_d\!=\!1)+\textrm{Li}^{3+}$ and is attractive due to polarization effects. This potential allows for bound states. Here $n_d$ and $n_{\rm Li}$ are the principal quantum numbers of a state of the muonic hydrogen-like deuterium atom and lithium ion, respectively. Other adiabatic states, which are below the $3d\sigma$, converge to the cluster: $d+\textrm{Li}^{3+}\mu(n_{\rm{Li}})$ with the principal quantum numbers $n_{\rm Li}=1,2$ for the muonic lithium ion and are repulsive. Thus we have four open channels below the $d\mu(n_d\!=\!1)$ threshold and the bound states associated with the $3d\sigma$ potential. Due to nonadiabatic coupling the adiabatic states become quasibound.

\section{Variational expansion and the CCR method}

The system of interest consists of three particles, a negative muon of mass $m_{\mu}$ and two nuclei of masses $M_a$ and $M_b$, where $a$, hereafter, stands for a lithium nucleus and $b$ for the deuteron. The Hamiltonian (in muonic atomic units: $|e|=\hbar=m_{\mu}=1$), after separating the center of mass motion, can be written as
\begin{equation}\label{Hamiltonian}
H = -\frac{1}{2\mu_1}\Delta_{{\mathbf{r}}_1}-\frac{1}{2\mu_2}\Delta_{{\mathbf{r}}_2}
   -\nabla_{{\mathbf{r}}_1}\cdot \nabla_{{\mathbf{r}}_2}
   -\frac{3}{r_1}-\frac{1}{r_2}+\frac{3}{R}
   \equiv T+V,
\end{equation}
where $\mathbf{r}_1$ and $\mathbf{r}_2$ are the vectors towards the muon from the two nuclei, $R$ denotes the distance between the nuclei, and $\mu_i=m_{\mu}M_i/(m_{\mu}+M_i)$ are the reduced masses of the respective muonic atoms ($i=1,2$).

In order to solve the three-body Coulomb problem for a quasibound state, we employ the Complex Coordinate Rotation (CCR) method which has been successfully used in atomic physics for studying properties of resonant states (see, for example, Ref.\cite{Reinhardt,Ho} and references therein). The advantage of this approach is that it has a rigorous mathematical background \cite{BalCom,Simon}.

The resonant state is defined as the solution of the eigenvalue problem
\begin{equation}\label{eigen}
(H(\theta)-E_{\rm res})\Psi_\theta = 0,
\qquad
H(\theta) = U(\theta)HU^{-1}(\theta),
\end{equation}
for the Hamiltonian $H(\theta)$ depended on dilation parameter $\theta$. This parameter scales all the coordinates of the system: $r_{ij}e^{\theta}$, and for many known interactions, $V(\theta)$ is an analytical function of $\theta$ and may be analytically continued into the complex plane \cite{ReedSimon4} (see Sec.\ XIII.10). Such a transformation has a great computational advantage for systems with Coulomb interactions \cite{Reinhardt,Ho}. The kinetic and potential parts transforms as $\exp{(-2\theta)}$ and $\exp{(-\theta)}$, respectively, and the Hamiltonian can be written as
\begin{equation}\label{DHamiltonian}
H(\theta) = Te^{-2\theta}+Ve^{-\theta}.
\end{equation}
The continuum spectrum of $H(\theta)$ is rotated on the complex plane around branch points ("thresholds") to "uncover" resonant poles situated on the unphysical sheet of the Reimann surface in accordance with the Aguilar-Balslev-Combes theorem \cite{BalCom}. The eigenfunction $\Psi_{\theta}$ obtained from Eq.~(\ref{eigen}), is square-integrable and the corresponding complex eigenvalue $E_{\rm res} = E_r - i\Gamma/2$ defines the energy $E_r$ and the width of the resonance, $\Gamma$, the latter is being related to the Auger rate as $\lambda_A = \Gamma/\hbar$.

In our numerical calculations we use the variational method based on exponential expansion with randomly generated
complex parameters. This approach has been developed in a variety of works \cite{Ros71,Smith,Frolov95}. Details and particular strategy of choice of the variational nonlinear parameters and construction of the basis sets that have been adopted in the present calculations can be found in \cite{Korobov00}.

Briefly, the wave function for a state with a total orbital angular momentum $L$ and of a total spatial parity $\pi=(-1)^L$ is expanded as follows ($\mathbf{r}_2 = \mathbf{r}_1\!-\!\mathbf{R}$):
\begin{equation}\label{exp_main}
\begin{array}{@{}l}
\displaystyle \Psi_{LM}^\pi(\mathbf{R},\mathbf{r}_1) =
       \sum_{l_1+l_2=L}
         \mathcal{Y}^{l_1l_2}_{LM}(\mathbf{R},\mathbf{r}_1)
         G^{L\pi}_{l_1l_2}(R,r_1,r_2),
\\[4mm]\displaystyle
G_{l_1l_2}^{L\pi}(R,r_1,r_2) =
    \sum_{k=1}^N \Big\{C_k\,\mbox{Re}
          \bigl[e^{-\alpha_k R-\beta_k r_1-\gamma_k r_2}\bigr]
+D_k\,\mbox{Im} \bigl[e^{-\alpha_k R-\beta_k r_1-\gamma_k r_2}\bigr] \Big\}.
\end{array}
\end{equation}
where $\mathcal{Y}^{l_1l_2}_{LM}(\mathbf{R},\mathbf{r}_1)$ are the solid bipolar harmonics defined as in Ref. \cite{Var88},
\[
\mathcal{Y}^{l_1l_2}_{LM}(\mathbf{R},\mathbf{r}_1) =
   R^{l_1}r_1^{l_2}\left\{Y_{l_1}\otimes Y_{l_2}\right\}_{LM},
\]
and the complex parameters, $\alpha$, $\beta$, $\gamma$, are generated in a pseudorandom manner \cite{Frolov95,var99}:
\begin{equation}\label{tempering}
\begin{array}{@{}l}\displaystyle
\alpha_k =
   \left[\left\lfloor\frac{1}{2}k(k+1)\sqrt{p_{\alpha}}\right\rfloor(A_2-A_1)+A_1\right]
\\[3mm]\displaystyle\hspace{10mm}
   +i\left[\left\lfloor\frac{1}{2}k(k+1)\sqrt{q_{\alpha}}\right\rfloor(A'_2-A'_1)+A'_1\right]\,,
\end{array}
\end{equation}
where $\lfloor{x}\rfloor$ designates the fractional part of $x$, $p_{\alpha}$ and $q_{\alpha}$ are some prime numbers, and $[A_1,A_2]$ and $[A'_1,A'_2]$ are real variational intervals, which need to be optimized. Parameters $\beta_k$ and $\gamma_k$ are obtained in a similar way. When exponents $\alpha_k$, $\beta_k$, and $\gamma_k$ are real, the method reveals slow convergence for molecular type Coulomb systems. Thus the use of complex exponents allows to reproduce the oscillatory behaviour of the vibrational part of the wave function and to improve convergence \cite{Frolov95,Korobov00}.

\begin{figure}[t]
\begin{center}
\includegraphics*[width=0.6\textwidth]{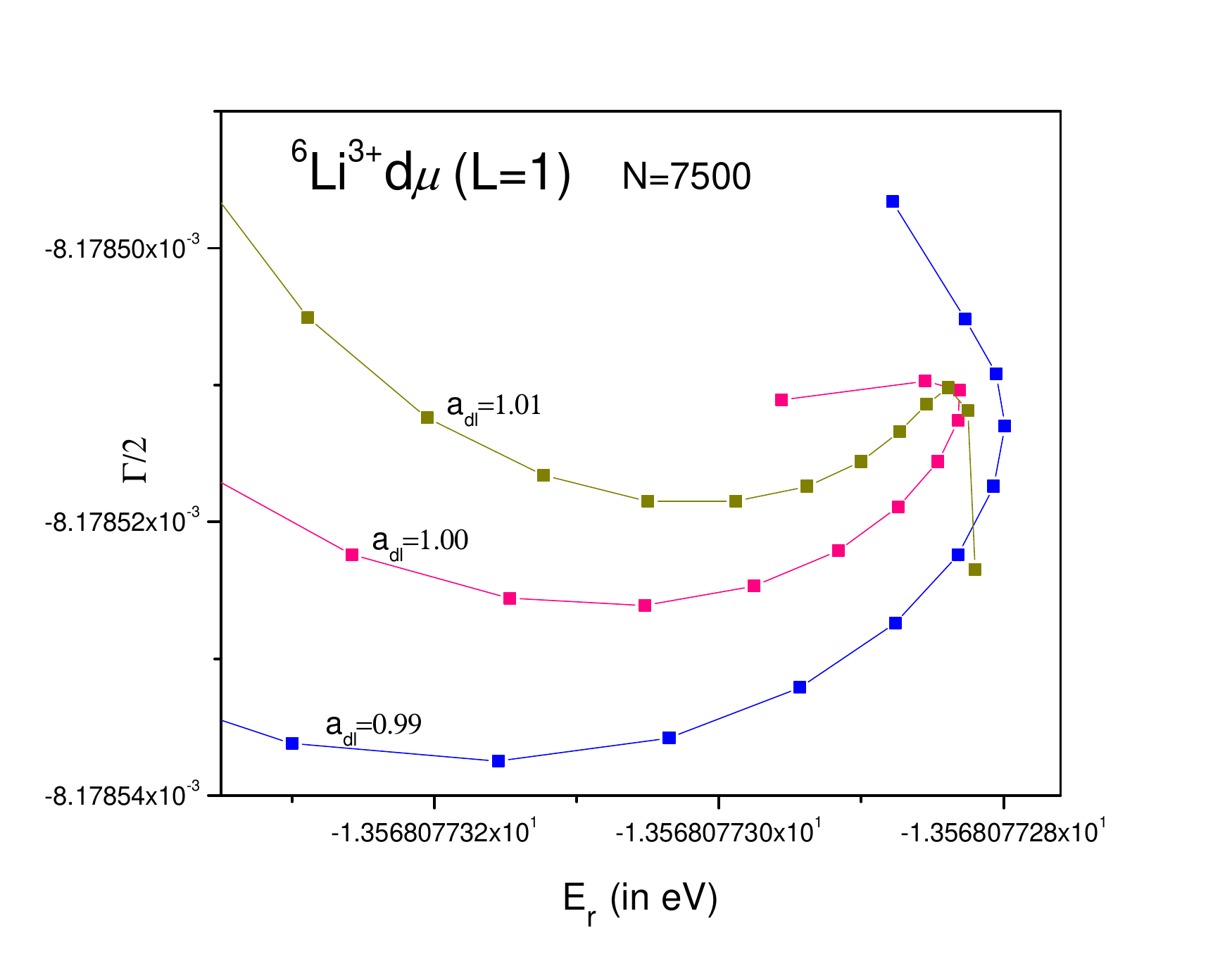}
\end{center}
\caption{Rotational paths, energy ($E_r$), and half-width ($\Gamma/2$) of the resonant state $^6$Li$^{3+}d\mu(L\!=\!1)$. The three paths with dilatation parameters ($a_{dl}=e^{\mathrm{Re}[\theta]}$): $a_{dl}=0.99,1.0,1.01$ and rotational parameters $\varphi=\mathrm{Im}[\theta]=0.10,0.11,\dots,0.30$ were used in calculations.}\label{fig:ccr}
\end{figure}

The exact eigenvalues do not depend on the rotating parameter $\theta$, still when one has a numerical approximation with the finite basis set one may expect only approximate eigenvalues which, therefore, have such dependence. In this case, the resonance positions and widths are deduced from the condition that a discrete complex eigenvalue is stabilized, i.e.
\begin{equation}
        \frac{\partial E_{\rm res}}{\partial\theta}=0,
\end{equation}
with respect to variations of the complex dilation parameter $\theta$ \cite{BraFro}.

\section{Results}

To solve Eq.~(\ref{eigen}) numerically, the inverse iteration method adapted to symmetric complex matrices has been employed. This method is very efficient in computational time and stable to round-off errors. To get the energy position and width of resonant states of interest we use trial wave functions with up to 7500 terms. An illustration of particular calculations for the $P$ state in $^6$Li$^{3+}d\mu$ ion is shown on Figure \ref{fig:ccr}. Taking the stationary point of the rotational paths one gets parameters of the resonance: $E_r = -13.56807729$ eV and $\Gamma/2 = 0.00817851$ eV. $E_r$ is reckoned from the $d\mu(n_d\!=\!1)+\,^6\textrm{Li}^{3+}$ threshold energy. Other results of our calculations are summarized in Table \ref{results}. For the states of $L=0,1$ the numerical precision obtained corresponds to a number of digits indicated in the Table, while in case of $L=2$ state we show in parentheses the error bars, which were obtained by studying the convergence. Error bars for the imaginary part are the same, since we define numerically an eigenvalue as a point on the complex plain.

\begin{table}[t]
\caption{Complex CCR energies: $E_{\rm res} = E_r+i\Gamma/2$, for the states of ${\,}^6$Li$^{3+}d\mu$ and $^7$Li$^{3+}d\mu$ ions. Here $E_r$ is the "binding" energy reckoned from the $d\mu(n\!=\!1)$ dissociation threshold and $\Gamma$ is the width of the resonance (in eV).}\label{results}
\begin{center}
\begin{tabular}{c@{\hspace{6mm}}c@{\hspace{6mm}}c@{\hspace{6mm}}c}
\hline\hline
 & $L=0$ & $L=1$ & $L=2$ \\
\hline
\vrule height11pt width0pt
$^6$Li$d\mu$ &
 $20.30807824+i\,0.00676920$ & $13.56807729+i\,0.00817851$ & $1.678490(9)+i\,0.007898$ \\
\vrule height11pt width0pt
$^7$Li$d\mu$ &
 $21.45595014+i\,0.00381025$ & $14.81736783+i\,0.00509516$ & $2.929779(3)+i\,0.005726$ \\
\hline\hline
\end{tabular}
\end{center}
\end{table}

For the nuclear fusion reactions, which may occur from these states, a key quantity is $|\Psi(0)|^2$, the squared amplitude of the wave function at the coalescence point of two nuclei. For the CCR formalism, the analytically dependent quantity is $\Psi^2(0)$,
\begin{equation}
\Psi^2_{\theta}(0) = \int
\left[\Psi_{\theta}(0,\mathbf{r})\right]^2\,d\mathbf{r}\,,
\end{equation}
and this quantity may be calculated in the same way as the energy: it becomes stationary when the variational wave function solution approaches the exact wave function of the system. In case of exact solution, $\Psi^2_{\theta}(0)$ is a constant function of rotational angle $\varphi=\mathrm{Im}[\theta]$ and may be extrapolated to $\varphi\to0$, to the real axis. In this way we get the following data for $\Psi^2(0)$ for the $S$-state in $^6$Li$^{3+}d\mu$ ion:
\[
\Psi^2(0) = [-3.0(1)-i1.2(1)]\times10^{-19}\mbox{ mau},
\]
and
\[
\Psi^2(0) = [-1.03(3)-i1.07(3)]\times10^{-19}\mbox{ mau},
\]
for the $S$-state in $^7$Li$^{3+}d\mu$ ion. The corresponding values for the physical quantity are:
\[
|\Psi(0)|^2=1.9(1)\times10^{-26}\hbox{ fm}^{-3},
\qquad
\hbox{for $^6$Li$\,d\mu$},
\]
and
\[
|\Psi(0)|^2=8.8(1)\times10^{-27}\hbox{ fm}^{-3},
\qquad
\hbox{for $^7$Li$\,d\mu$}.
\]
For the states with total orbital angular momentum $L=1,2$ the fusion comes mainly via the $P$-wave mode and is strongly suppressed. In these calculations in order to get convergent results we had to increase the basis set to $N=20\>000$ and use multiprecision arithmetics with one hundred of decimal digits.

The values of masses adopted in our calculations: $M_{{}^6{\rm Li}^{3+}}=10961.8982545\>m_e$, $M_{{}^7{\rm Li}^{3+}}=12786.391884\>m_e$, $M_d=3670.48296785\>m_e$ and $m_{\mu}=206.7682826\>m_e$. The Rydberg constant as a conversion coefficient to eV was taken: $hcR_\infty=13.605\,693\,009$ eV.

\section{Conclusion}

In conclusion, we want to state that we found six quasibound states (three states for each isotope) supported by the $3d\sigma$ adiabatic potential in the lithium deuteride muonic molecular ions, and it is clearly seen from Table 1 that the higher rotational states should be unbound. For the first time, the reliable estimates for the probability density of the wave function at $R\!=\!0$, were obtained with the precision of about 3\% or even better. This is one of the key quantities for experimental studying of the fusion rate for $d+\hbox{Li}$ nuclear reactions. It may allow to measure in the experiment the astrophysical $S$-factor $S(E)$ \cite{Thompson}:
\[
S(E) = \frac{E}{\exp(-2\pi\eta)}\sigma(E),
\qquad
\eta = \frac{Z_1Z_2e^2}{4\pi\epsilon_0\hbar v},
\]
for the low-energy scattering of nuclei. Here $\sigma(E)$ is the total cross section of the fusion reaction, $v$ is the relative incident velocity. $S(E)$ allows to account for the Coulomb repulsion between charged nuclei. Thus $S(E)$ is a slowly varying function when $E$ tends to zero.

We have to acknowledge that our previous attempt to solve this problem \cite{Rakit96} using the variational CCR calculations had failed due to less capable computers and, as a consequence, much poor convergence of the expansion.

\section*{Acknowledgements}

Support from the Russian Science Foundation under Grant No.~18-12-00128 is gratefully acknowledged.

\end{document}